\documentclass[12pt]{iopart}
\usepackage{graphicx}
\graphicspath{{figures/}{figures/converted/}}

\usepackage{iopams}
\newcommand{\text}[1]{\ensuremath{\mathrm{#1}}}

\usepackage[numbers, sort&compress]{natbib}
\newcommand{\newblock}{} %required by natbib

\begin{document}

\providecommand{\ignore}[1]{}
%\newcommand{\comment}[1]{}
% Comment the following to turn off comments:
%\renewcommand{\comment}[1]{{\color[rgb]{0,1,0}[#1]}}
\newcommand{\ket}[1]{{{|}{#1}\rangle}}
\newcommand{\bra}[1]{{\langle{#1}{|}}}
\newcommand{\braket}[2]{{\langle{#1}{|}{#2}\rangle}}

\title{Quantum state preparation and control of single molecular ions}
\author{D. Leibfried}
\address{Time and Frequency Division, National Institute of Standards and Technology, 325 Broadway, Boulder, CO 80305, USA}
\begin{abstract}
Preparing molecules at rest and in a highly pure quantum state is a long standing dream in chemistry and physics, so far achieved only for a select set of molecules in  dedicated experimental setups. Here, a quantum-limited combination of mass spectrometry and Raman spectroscopy is proposed that should be applicable to a wide range of molecular ions. Excitation of electrons in the molecule followed by uncontrolled decay and branching into several lower energy states is avoided. Instead, the molecule is always confined to rotational states within the electronic and vibrational ground-state manifold, while a co-trapped atomic ion provides efficient entropy removal and allows for extraction of information on the molecule. The outlined techniques might enable preparation, manipulation and measurement of a large multitude of molecular ion species with the same instrument, with applications including, but not limited to, precise determination of molecular properties and fundamental tests of physics.
\end{abstract}

\section{Introduction}

In the last 30 years, laser cooling\cite{han75,win75} and laser spectroscopy has led to spectacular developments in atomic physics. The electromagnetic vacuum of infrared to ultraviolet light provides a near perfect entropy sink and allows us to optically pump and cool atoms, if sufficiently closed cycles of electronic excitation and subsequent decay are realized. Entropy can be removed from the electronic state and the translational motion to where these degrees of freedom are in a highly pure quantum state. Such quantum state preparation is available for a sizeable number of neutral and charged atoms, but in most molecules the decay of excited electrons branches into a multitude of levels, and this prevents efficient entropy removal in all but a few cases. These exceptions include cooling molecules by buffer gas\cite{wei98}, exploiting fortuitous level structures that allow for closed cycles with a manageable number of laser sources\cite{shu10,sta10,sch10}, selective ionization and charge transfer\cite{ton10} or assembling molecules from ultracold atoms\cite{ni08,dei08,dan08,lan08}. In all of these cases, the molecular states are destroyed, or the molecules leave the interaction region during the process of their detection. Methods to prepare ensembles of specific molecules in their absolute ground state by pulsed adiabatic passage\cite{pee07, sha08,laz10} or by collisions with a laser cooled buffer gas\cite{hud09}  have also been envisioned.\\
% leave trap/region: doyle, demille,
% dest: drewsen, schiller, ni, danzl, deiglmayr, lang, tong
% theory: Shapiro, Je&Co.
\section{Translational cooling of molecular ions}
In the method proposed here, the molecular ion of interest is co-trapped with an atomic ion that can be precisely manipulated by laser and microwave fields\cite{lei03}. Due to the Coulomb-coupling between the co-trapped ions, laser-cooling of the atomic ion will sympathetically cool the translational motion of the molecular ion as well. Typically, excited electronic and vibrational states of the molecular ion will decay to their ground states within nanoseconds and milliseconds, respectively, and the rotational state distribution will equilibrate to the black-body temperature of the trap within a few seconds\cite{sta10,sch10}. After this initial equilibration, the molecule ideally never leaves the electronic and vibrational ground-state manifold. The mass of the molecular ion can be inferred quite precisely by determining the normal-mode frequencies of motion of the co-trapped ions\cite{win83,dre04}. This can be accomplished by sweeping the frequency of an electronic drive signal that is applied to a trap electrode\cite{win83}. If the drive frequency coincides with a normal-mode frequency, the crystal gets excited and the laser-induced fluorescence changes. If the mass of the atomic ion $m_a$ is known and it has frequency $\omega_a$ in the potential well, the two-ion crystal with mass ratio $\mu=m_m/m_a \geq 1$ will exhibit two axial modes, with frequencies \cite{mor01}
\begin{equation}\label{Eq:NorFre}
\omega_\pm = \omega_a \sqrt{ \left(1+\frac{1}{\mu } \pm \sqrt{1+\frac{1}{\mu ^2}-\frac{1}{\mu }}\right) },
\end{equation}
and eigenvectors
\begin{equation}
v_\pm = \frac{1}{\sqrt{\left(1-\mu \mp \sqrt{\mu ^2-\mu +1}\right)^2+1}}
 \left\{1-\mu\mp \sqrt{\mu ^2-\mu +1},1\right\},
  \end{equation}
where the first coordinate is of the the lighter mass atomic ion. If the molecule mass $m_m$ is smaller than $m_a$ similar relations can be derived. From the measured frequencies $\omega_a$ and $\omega_\pm$, which can be determined with better than $10^{-4}$ relative accuracy\cite{dre04}, we can infer $m_m$:
\begin{equation}
m_m =m_a \frac{2 \omega_a^2}{\omega_+^2 +\omega_-^2-2 \omega_a^2}.
\end{equation}
For $\mu \gg 1$, $\omega_+$  goes to $\sqrt{2}~ \omega_a$ while $\omega_-$ goes to zero. In the same limit the normal-mode amplitudes of the ions go to $\{-1,0\}$ in the $\omega_+$- mode and to $\{1/\sqrt{5},2/\sqrt{5}\}$ for the $\omega_-$-mode. Cooling of both modes will become more of a challenge with larger mass differences: For the $\omega_+$-mode, the participation of the heavier ion tends to vanish, while the $\omega_-$-mode will go to low frequencies, where resolving sidebands becomes harder and ambient heating mechanisms are more detrimental\cite{lei03}.

For not too different masses, it should be possible to cool the normal modes of motion of both ions to the ground state, analogous to ground state cooling demonstrated with different species atomic ions\cite{bar03,sch05} which has been implemented for a range of masses $1/3 \leq m_a/m_{a'} \leq 2$. Outside this range, ground-state-cooling becomes less effective, but might still be feasible. Suitable atomic ions range from 9 atomic mass units ($^9$Be$^+$) to 176 atomic mass units ($^{176}$Yb$^+$). Only one of the normal modes is required to be in the ground state for what follows. At this point the molecule is ideally in the electronic, vibrational and translational ground states, leaving only the rotational and, if the molecule has hyperfine structure, the hyperfine states undetermined.

\section{Rotational spectroscopy and cooling}

\begin{figure}[htbp]
\begin{center}
\includegraphics[width=13 cm,type=pdf,ext=.pdf,read=.pdf]{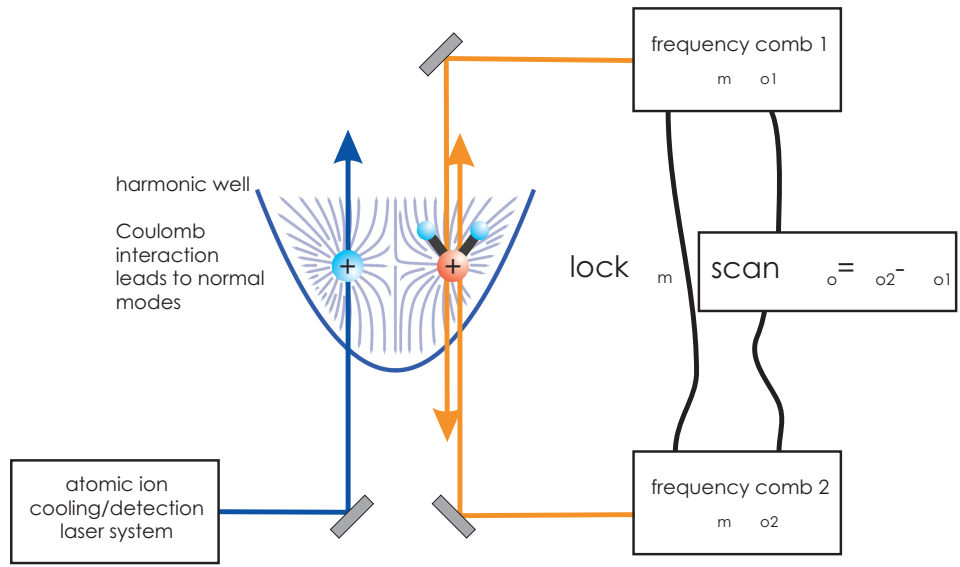}
\caption{Schematic setup of a quantum-limited molecular spectrometer. A known atomic ion (blue) and a, possibly unknown, molecular ion (red and blue stick model) are confined in the same harmonic trapping well. The atomic ion is initialized to a known internal state. Due to the Coulomb interaction between atomic and molecular ions the normal modes of vibration of the entire crystal can be sympathetically cooled to their ground state by coupling exclusively to the atomic ion with laser interactions (blue arrow). The rotational states of the molecule are then manipulated by inducing Raman-transitions due to the light fields of two counter-propagating frequency combs (orange arrows) of equal repetition rate $\omega_m/(2 \pi)$ and variable relative carrier-envelope-phase $\Delta \omega_o$. The photon recoils produced by Raman-transitions of the molecule can excite the shared crystal motion, and this excitation can subsequently be detected on the atomic ion (see text). Detection of crystal excitation thus establishes that the molecule has undergone a transition.}
\label{Fig:SchSet}
\end{center}
\end{figure}

Manipulation of rotational states, ideally without altering the vibrational and the electronic state, might proceed by combining Raman-transitions\cite{ram28} driven by optical frequency combs\cite{ye05} with a technique similar to what has been used for state preparation and read-out in two-ion quantum logic clocks\cite{sch05}. In this way, entropy can be removed from the molecular rotational degree of freedom by indirectly coupling it to the near perfect reservoir of a light field. This can be accomplished via closed-cycle electronic transitions of the co-trapped atomic ion. Raman transitions are a near-universal tool in molecular spectroscopy, because they can be driven with light sources that are not specific to the molecule under study. Direct electronic transitions in molecular ions typically require wavelengths shorter than 650 nm\cite{wil80}, below the center wavelengths of common femtosecond laser frequency combs (mostly 800 nm or 1064 nm)\cite{ye05}. Two comb beams can be produced either by splitting the output of one laser and frequency shifting one part\cite{hay10}, or by using two separate combs with equal mode spacing and scanning the carrier-envelope phase of one or both. If the combs span about 30 THz, they are much broader than the transition frequencies between the highest lying thermally occupied rotational states that need to be addressed (typically below 300 K $k_B/h~\approx$  6.25 THz  where $k_B$ and $h$ are Boltzmann's and Planck's constant respectively\cite{wil80}). The two comb fields should counter-propagate along the ion crystal axis for maximal momentum transfer in transitions that change the state of one normal mode of the ion crystal while simultaneously driving a rotational state changing transition in the molecule (see Fig.\ref{Fig:SchSet}).
Analogous processes have been utilized in experiments with a laser comb and two identical atomic ions where an approximately 12.7 GHz hyperfine Raman transition was driven simultaneously and coherently with the ion crystal motion\cite{hay10}. Raman transitions of the molecule require that the resonance condition
\begin{equation}\label{Eq:ResCon}
N \omega_m+\Delta \omega_o = \Omega_{J,J\pm1,2} + n~\omega_\pm
\end{equation}
is fulfilled. Here $N$ and $n$ are integers, $\omega_m$ is the mode spacing of both comb fields, set to be identical, and $\Delta \omega_o$ is the frequency shift between the two comb fields due to their individual carrier-envelope phases (or due to the frequency shift of one part of the output). The frequency difference  between two rotational levels with quantum numbers $J$ and $J\pm1,2$ is denoted  $\Omega_{J,J\pm1,2}$  (for linear and spherical-top-molecules, $J\rightarrow J\pm1$ is forbidden by Raman selection rules) and  $\omega_\pm$ is either one of the normal mode frequencies of the two-ion crystal.  Transitions with $n=0$ (``carrier'') will have the largest transition rate, while the $n$-th ``sideband'' will be suppressed by the $n$-th power of the Lamb-Dicke parameter. For Raman-transitions induced by two counter-propagating beams with wave-vector modulus $k_l$, the Lamb-Dicke parameters of the molecular ion are given by\cite{mor01}
\begin{equation}\label{Eq:LamDic}
\eta_m^{(\pm)} = 2 |k_l| v_{\pm}^{(2)} \sqrt{\frac{\hbar}{2 m_m \omega_\pm}}.
\end{equation}
At mode frequencies commonly used in resolved sideband cooling\cite{lei03}, the ground state wavefunction extension of the molecule in either normal mode (typically a few nanometers) is much smaller than the light wavelength $\lambda$, so higher-order sidebands are largely suppressed. The first sidebands ($n=\pm 1$) are the strongest motion-altering interaction and change the motional state by one quantum.\\

For a very general rough estimate of achievable Raman rates, we can assume that the intermediate level in the Raman transition has a dipole matrix element of $e\cdot a_0 \approx$  8.48 10$^{-30}$ C$\cdot$m, where $e$ is the electron charge and $a_0$  the Bohr radius. The transition rate on the $n=0$ carrier is then approximately\cite{win03}
\begin{equation}\label{Eq:RamEst}
|\Omega_0| =\frac{(e~ a_0 E_{\rm av})^2}{4 \hbar \Delta_l},
\end{equation}
with $\Delta_l$ the detuning of the comb center frequency to the resonance frequency from the molecular ground state to the intermediate level and $E_{\rm av}$ the average electric field per comb, set to be identical for both combs for simplicity. If each comb emits 1 W on average into a beamwaist of 20 $\mu$m around 800 nm and the dipole transition resonance is at 400 nm, the $n=0$ carrier Raman transition rate is $|\Omega_0|/(2 \pi) \approx$ 824 kHz. For the $n=\pm1$ sideband transitions this frequency must be multiplied by the Lamb-Dicke parameter $\eta$, which typically ranges between 0.03 and 0.3, depending on the normal mode structure and the comb wavelength [see Eq.(\ref{Eq:LamDic})].\\

\subsection{Rotational cooling for known molecular constants}
\label{Sec:RotCoo}

\begin{figure}[htbp]
\begin{center}
\includegraphics[width=16 cm,type=pdf,ext=.pdf,read=.pdf]{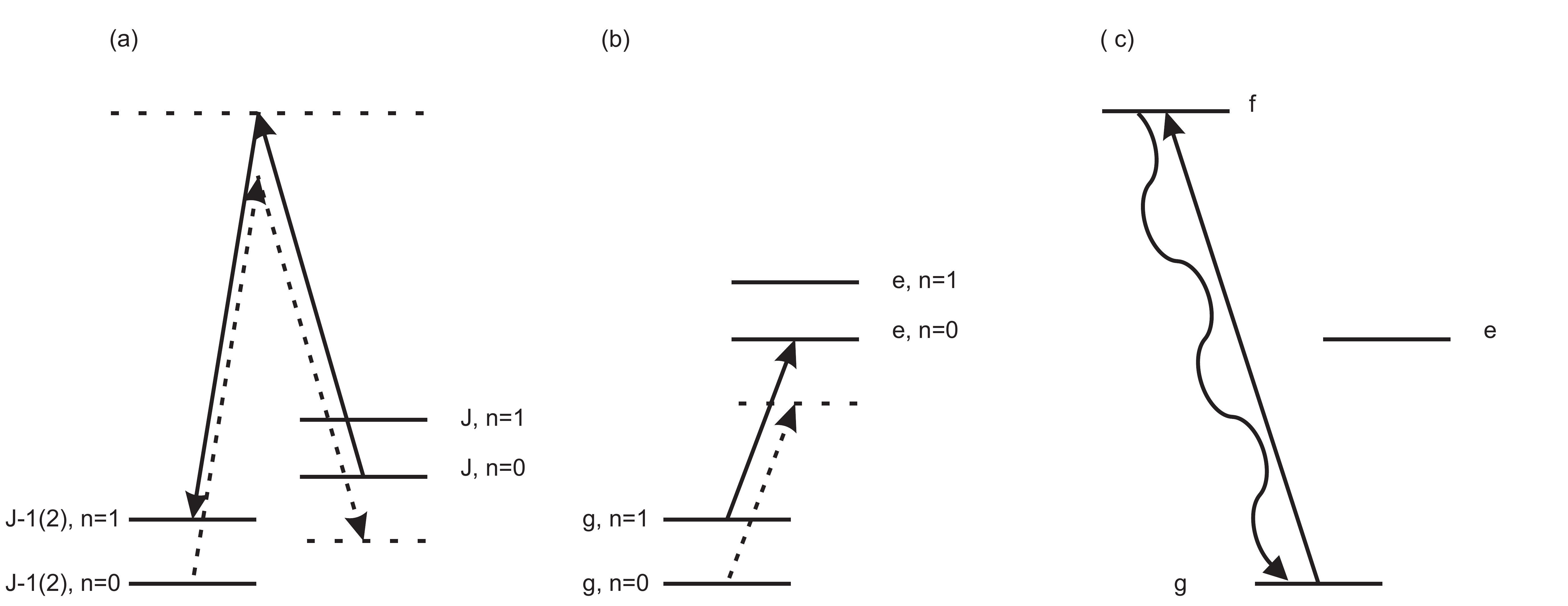}
\caption{Molecular rotational Raman transition and subsequent detection (energy differences not to scale). (a) The molecule undergoes a Raman-transition on the $n=-1$ sideband of the selected normal mode of the two-ion crystal in Eq.(\ref{Eq:ResCon}). Since this mode is in the ground state, only the rotational transition $J \rightarrow J-1$ is energetically allowed (solid lines) while the opposite direction is not resonant with an existing state (dashed lines). (b) The conditionally added quantum of motion is transferred into an excitation of the co-trapped atom by use of a $(g,n=1) \rightarrow (e,n=0)$ resolved sideband transition, where $g$ and $e$ refer to two long-lived states in the atomic ion. (c) The electronic state of the atomic ion is detected by laser induced fluorescence on a cycling transition connecting $g$ with the fast-decaying level $f$\cite{sch05,lei03}. If no fluorescence is detected, the atom is found to be in excited state $e$. This near-ideal measurement projects the state of the molecule into $J-1$. The atom can then be reset to $g$ by optical pumping and the crystal can be re-cooled to the motional ground state. This resets atom and two-ion-crystal motion to the initial condition before step (a) without affecting the rovibrational or electronic state of the molecule. The reset removes entropy from the system and prepares it for further manipulations.}
\label{Fig:RamTra}
\end{center}
\end{figure}

For known rotational constants and transition rates of the molecule, we can set the relative detuning of the combs such that two rotational levels with high thermal occupation probability are connected by a $n=-1$ resonance condition in Eq.(\ref{Eq:ResCon}). As long as the motion of the two-ion-crystal is in the ground state, this transition can proceed only one way, from the $J$ to the $J-1,2$ states, while one quantum of motion is added to the mode at $\omega_\pm$ (see Fig. \ref{Fig:RamTra}(a)), the reverse direction cannot conserve energy.
Each attempt to excite a Raman transition between rotational states of the molecule with the comb-lasers would be followed by an attempt to drive a $n \rightarrow (n-1)$ resolved sideband on the cooling and detection ion (see Fig. \ref{Fig:RamTra}(b)). As long as the selected normal mode of crystal motion is in the ground state ($n=0$), this transition is energetically forbidden, but if the preceding transition of the molecule has deposited a quantum of motion ($n=1$), the electronic state of the atomic ion can be changed while this quantum is removed~\cite{sch05}. The electronic state-change of the atom can be detected by state-selective fluorescence on the cooling and detection transition, in analogy to quantum-logic ion clocks~\cite{sch05,lei03} (see Fig. \ref{Fig:RamTra}(c)). Crucially, if a state change of the atomic ion is detected, this near-ideal measurement projects the rotational state of the molecule into a pure and known quantum state, reducing the entropy of the molecular state considerably in one successful attempt\cite{vog06}. In case the excitation attempt is not successful, the addressed rotational state is almost certainly not occupied, so we can proceed to probing another highly occupied rotational state in an updated distribution that excludes the unsuccessfully probed level(s). Once the initial distribution of rotational states is projected into a pure state in this manner, we can drive the system straight into the rotational ground state with carrier-type transitions ($n=0$ in Eq.(\ref{Eq:ResCon})), if the Raman transition rates are known. It might also be possible to use techniques similar to stimulated rapid adiabatic passage (STIRAP)\cite{ber98,sha08,laz10} to provide more robust carrier and sideband transitions without exact knowledge of the transition rates.

\subsection{Determination of transition rates}
\label{Sec:DetTra}

If the Raman transition rates of the molecule are not known, they can be determined as follows: Once the molecular ion has been projected into rotational state $J$, we can probe the molecule on the $(J,n=0) \rightarrow (J-1,2, n=1)$ transition for several different excitation periods and transfer any motional excitation of the two-ion crystal into an electronic state change of the atom by inducing a red sideband transition of the atom, as in the preceding step (section \ref{Sec:RotCoo}) that narrowed the thermal $J$-distribution to just one state. If the atomic ion changes state in such an excitation attempt, the transition to $(J-1,2, n=1)$ has most likely been made; therefore, the molecule is now in state $J-1,2$ with high probability, and can be returned to $J$ by re-cooling the two-ion crystal to the $n=0$ ground state, followed by repeated attempts to do a $(J-1,2,n=0) \rightarrow (J,n=1)$ sideband transition (detuned by 2 $\omega_\pm$ to the blue from the transition previously attempted). By recording transition probabilities as a function of length of excitation in many repetitions of this cycle, the transition rate and subsequently the magnitude of transition matrix elements can be inferred.

\subsection{Quantum-non-demolition measurement of the rotational state}
\label{Sec:QuaNon}

For known transition rates, one should be able to repeat the cycle of successive $n=\pm1$ transitions and detection of the molecular state-changes outlined in section \ref{Sec:DetTra} several times. This ideally constitutes a quantum-non-demolition (QND) readout as we should leave a motional quantum that can be subsequently read out through the state of the atom in each cycle step. Every time this happens, we gain more certainty about the state the molecule is in, even if single measurements have only modest fidelity.  An analogous state-detection scheme with atomic ions has reached a detection fidelity  of 99.94~\%, with a  fidelity of about 80~\% for a single measurement \cite{hum07}. Several cycles of such QND readouts also prepare the molecule rotation into a pure quantum state with higher confidence than that for a single detection of a transition.

\subsection{Determination of molecular constants}
\label{Sec:DetRot}

If the rotational levels of the molecular ion are unknown, we can slowly scan the frequency difference $\Delta \omega_o$ of the two combs over the comb spacing $\omega_m$. At some point, a $n=\pm 1$ resonance will lead to a molecular Raman-transition that will leave a quantum in the selected mode of motion of the crystal. Detecting a state change in the atom after converting the motion into atomic excitation implies that the molecule is in either one of two $J$-levels that the previous Raman-interaction has resonantly connected, so the distribution of rotational states is narrowed to two levels. We can reset the atom state by optical pumping, and recool the crystal motion to the ground state, and then attempt to drive a molecular transition at a relative detuning that is 2 $\omega_\pm$ red-detuned from the previous pulse. Again, the transition is possible only if a quantum of motion is added that can be subsequently detected with the atom.  Detection of the atomic state change would establish that the molecule has undergone a transition from the higher energy $J$ to the $J-1$ or $J-2$ state (for linear and spherical-top molecules, the $J-2$ state has been prepared with certainty). If several attempts to drive at this red-detuned frequency fail, the molecule is most likely already in the lower-energy state and the failed attempts were actually on a $n=-3$ sideband transition that is also energetically forbidden if the selected mode of crystal motion is in the ground state. In this case, attempts to drive  transitions 2 $\omega_\pm$ blue detuned from the pulse that produced the original transition should excite the molecule to the higher-energy $J$ state, while a motional quantum is deposited ($n=1$) that can subsequently be converted to a state change of the atom and detected. In either case, the unknown frequency $\Omega_{J,J\pm1,2}$ can be found from the mean of the two resonant frequencies in the preceding steps. It is then possible to find the rate of these transitions and gain information about the related transition matrix elements as described in section \ref{Sec:DetTra}. Further rotational levels may be identified by completing the scan of $\Delta \omega_o$ over the full comb mode spacing $\omega_m$. After several distinct transitions are found, it should be possible to assign $J$ and determine all rotational constants and centrifugal distortion constants. This can be done unambiguously, if the absolute frequency of every transition is known. This is possible even if $N$ in Eq.(\ref{Eq:ResCon}) is initially unknown. From the range of rotational transitions and assuming that $\omega_m$ is approximately $2 \pi \times 1$ GHz, $N$ can take values between 1 and $10^4$. After finding a resonance, we can change $\omega_m$ by a small amount of (say) order $10^{-3} \omega_m$, easily observable by monitoring the repetition rate of the comb(s) on a fast photodiode. Then $\Delta \omega_o$ can be scanned until the resonance condition is re-established. The closest integer to the fraction of the changes in $\omega_m$ and $\Delta \omega_o$ is $N$. We can eventually infer the full rotational spectrum of the molecule by systematically scanning the complete range $\omega_m$ and then determining $N$ for each resonance found. With this knowledge, the spectrum, originally  compressed into one range $\omega_m$, can be deconvoluted to yield rotational states information with a resolution limited only by the lifetimes of the involved states.\\

For simplicity, the preceding discussion has neglected any hyperfine structure that the molecule might possess. Hyperfine structure complicates the spectrum-finding procedure by adding more resonances and one more degree of freedom that needs to be determined. Still, it seems straightforward to incorporate hyperfine structure with slight modifications to the outlined procedures.

\section{MgH$^+$ co-trapped with Mg$^+$}

Estimated parameters for the concrete case of a MgH$^+$ molecule, co-trapped with a Mg$^+$ atom establish the feasibility of the proposed scheme. The rotational state of MgH$^+$ has been previously prepared close to its ground state in an ensemble of molecular ions with dedicated laser systems and a destructive detection method~\cite{sta10}; therefore all relevant molecular parameters are reasonably well known. As a linear molecule, MgH$^+$ admits only $J \rightarrow J\pm2$ rotational Raman transitions. The low mass and small size produce a rotational constant of approximately 190 GHz and a centrifugal correction of order 10 MHz. In thermal equilibrium with a room-temperature environment only the first 11 rotational states have occupation probabilities above 1~\%. The highest occupation probability is approximately 14~\% for $J=4$. The $J \rightarrow J\pm2$ selection rule dictates that we should first try to prepare a state with even $J$ to be able to connect to the ground state. The dipole transition matrix element between the ground X$^1 \Sigma^+$ and the lowest electronically excited A$^1  \Sigma^+$ state with an energy difference corresponding to 280 nm is about 1.57 $e~ a_0$\cite{jor05}. For the same laser parameters as in section \ref{Sec:RotCoo}, the carrier Raman transition rate, Eq.(\ref{Eq:RamEst}), is approximately 1.1 MHz. For a crystal consisting of MgH$^+$ and Mg$^+$ at a single Mg$^+$ trap frequency of $\omega_a=$1 MHz the two normal-mode frequencies, Eq. (\ref{Eq:NorFre}), are $\omega_+ \approx$ 0.99 MHz and $\omega_- \approx$ 1.72 MHz, with Lamb-Dicke factors, Eq. (\ref{Eq:LamDic}), of  $\eta_+ \approx$ 0.16 and $\eta_- \approx$ 0.13 respectively. These numbers imply $n=\pm 1$ sideband transition rates of 174 kHz (complete rotation between states achieved in 1.43 $\mu$s) and 137 kHz (complete rotation in 1.82 $\mu$s), respectively. Together with initial ground-state cooling and subsequent transfer of motional excitation and state readout of the Mg$^+$ ion, one experimental cycle should take no more than 1 ms\cite{lei03}. An initial scan could involve 100 attempts per comb detuning $\Delta \omega_o$,  yielding a scan speed of 10 frequency points per second. A 10 $\mu$s window with about 10000 Raman pulses applied at 1 GHz repetition rate to the molecule would correspond to a Fourier limited width of about 10 kHz for each resonance; so to collect a reasonably densely sampled spectrum over 1 GHz, we would need 10$^5$ points taken in 10$^4$ seconds, or about 3 hours. This duration represents a worst-case estimate, where nothing about the rotational constants is known, and is at the same time compatible with achievable lifetimes in ion traps. If we assume that 10 transitions with initial state population larger than 1~\% are present, it should take less than 1 hour to identify the first resonance. With each newly identified resonance, finding the remaining resonances should take fewer attempts. In this fashion, it should be possible to determine a complete rotational spectrum in a few hours. Once the most populated sub-levels are determined, these could be directly prepared within a time span comparable to the rotational re-thermalization period of the molecule by repeating excitation attempts until a transition is successfully made. An experiment in a low temperature environment would reduce the initial spread of $J$-levels considerably, while suppressing background gas collisions with the crystal and thus further increasing the ion lifetimes in the trap.\\

\section{Conclusions}

In the recent past, quantum control of neutral and ionized atoms has proven enormously fruitful. Controlling the quantum state of a multitude of molecules by use of the methods proposed here and also other methods that have recently emerged promises to take us to the next level of complexity. We can hope to keep and study the same molecule with a resolution limited only by natural linewidths under full control of its quantum states.
It should also be possible to efficiently prepare certain rotational states, including the rotational ground state and coherent superpositions of these states. This may be further aided by a higher fidelity, quantum-non-demolition way of measurement/preparation that is described in section \ref{Sec:QuaNon}. After preparation, the molecular quantum  state can be coherently manipulated with the same or other laser sources within the limits imposed by decay and re-thermalization of the involved states. For rotational states within the electronic and vibrational ground state manifold, re-thermalization is often on a timescale of many seconds.\\

Aided by these tools, we might test quantum mechanics and its application to coupled systems to unprecedented levels and search for energy shifts due to parity violation, a finite value of the electron electric dipole moment, or a variation of fundamental constants, such as the proton-to-electron mass ratio. It might also become feasible to induce highly controlled molecular collisions. Because reaction cross-sections can strongly depend on the vibrational and rotational quantum state, we might glean better control and understanding of chemical reactions in this way.\\

During completion of this manuscript I became aware of similar work in the group of D. Matsukevich at the National University of Singapore\cite{din11}.

\section*{Acknowledgments}

I thank Tobias Sch\"atz, Scott Diddams, Jun Ye and David Wineland for discussions and comments on the manuscript. Contribution of NIST, not subject to U.~S.~copyright.


\begin{thebibliography}{10}

\bibitem{han75}
T.~W. H\"{a}nsch and A.~L. Schawlow.
\newblock Cooling of gases by laser radiation.
\newblock {\em Opt. Comm.}, 13:68--69, 1975.

\bibitem{win75}
D.~J. Wineland and H.~Dehmelt.
\newblock Proposed {10$^{14}$ $\Delta \nu<\nu$} laser fluorescence spectroscopy
  on {Tl$^+$} mono-ion oscillator iii.
\newblock {\em Bull. Am. Phys. Soc.}, 20:637, 1975.

\bibitem{wei98}
Jonathan~D. Weinstein, Robert deCarvalho, Thierry Guillet, Bretislav Friedrich,
  and John~M. Doyle.
\newblock Magnetic trapping of calcium monohydride molecules at millikelvin
  temperatures.
\newblock {\em Nature}, 395:148--150, 1998.

\bibitem{shu10}
E.~S. Shuman, E.~S. Barry, and D.~DeMille.
\newblock Laser cooling of a diatomic molecule.
\newblock {\em Nature}, 467:820--823, 2010.

\bibitem{sta10}
Peter~F. Staanum, Klaus H{\o}jbjerre, Peter~S. Skyt, Anders~K. Hansen, and
  Michael Drewsen.
\newblock Rotational laser cooling of vibrationally and translationally cold
  molecular ions.
\newblock {\em Nature Phys.}, 6:1--4, 2010.

\bibitem{sch10}
T.~Schneider, B.~Roth, H.~Duncker, I.~Ernsting, and S.~Schiller.
\newblock All-optical preparation of molecular ions in the rovibrational ground
  state.
\newblock {\em Nature Phys.}, 6:275--278, 2010.

\bibitem{ton10}
Xin Tong, Alexander~H. Winney, and Stefan Willitsch.
\newblock Sympathetic cooling of molecular ions in selected rotational and
  vibrational states produced by threshold photoionization.

\bibitem{ni08}
K.-K. Ni, S.~Ospelkaus, M.~H.~G. de~Miranda, A.~Pe'er, B.~Neyenhuis, J.~J.
  Zirbel, S.~Kotochigova, P.~S. Julienne, D.~S. Jin, and J.~Ye.
\newblock A high phase-space-density gas of polar molecules.
\newblock {\em Science}, 322(5899):231--235, 2008.

\bibitem{dei08}
J.~Deiglmayr, A.~Grochola, M.~Repp, K.~M\"ortlbauer, C.~Gl\"uck, J.~Lange,
  O.~Dulieu, R.~Wester, and M.~Weidem\"uller.
\newblock Formation of ultracold polar molecules in the rovibrational ground
  state.
\newblock {\em Phys. Rev. Lett.}, 101(13):133004, Sep 2008.

\bibitem{dan08}
Johann~G. Danzl, Elmar Haller, Mattias Gustavsson, Manfred~J. Mark, Russell
  Hart, Nadia Bouloufa, Olivier Dulieu, Helmut Ritsch, and Hanns-Christoph
  N\"agerl.
\newblock Quantum gas of deeply bound ground state molecules.
\newblock {\em Science}, 321(5892):1062--1066, 2008.

\bibitem{lan08}
F.~Lang, K.~Winkler, C.~Strauss, R.~Grimm, and J.~Hecker-Denschlag.
\newblock Ultracold triplet molecules in the rovibrational ground state.
\newblock {\em Phys. Rev. Lett.}, 101:133005, 2008.

\bibitem{pee07}
Avi Pe'er, Evgeny~A. Shapiro, Matthew~C. Stowe, Moshe Shapiro, and Jun Ye.
\newblock Precise control of molecular dynamics with a femtosecond frequency
  comb.
\newblock {\em Phys. Rev. Lett.}, 98(11):113004, Mar 2007.

\bibitem{sha08}
Evgeny~A. Shapiro, Avi Pe'er, Jun Ye, and Moshe Shapiro.
\newblock Piecewise adiabatic population transfer in a molecule via a wave
  packet.
\newblock {\em Phys. Rev. Lett.}, 101(2):023601, Jul 2008.

\bibitem{laz10}
C.~Lazarou, M.~Keller, and B.~M. Garraway.
\newblock Molecular heat pump for rotational states.
\newblock {\em Phys. Rev. A}, 81:013418, Jan 2010.

\bibitem{hud09}
Eric~R. Hudson.
\newblock Method for producing ultracold molecular ions.
\newblock {\em Phys. Rev. A}, 79:032716, Mar 2009.

\bibitem{lei03}
D.~Leibfried, R.~Blatt, C.~Monroe, and D.~J. Wineland.
\newblock Quantum dynamics of single trapped ions.
\newblock {\em Rev. Mod. Phys.}, 75:281, 2003.

\bibitem{win83}
D.~J. Wineland, J.~J. Bollinger, and W.~M. Itano.
\newblock Laser fluorescence mass spectroscopy.
\newblock {\em Phys. Rev. Lett.}, 50:628--631, 1983.

\bibitem{dre04}
M.~Drewsen, A.~Mortensen, R.~Martinussen, P.~Staanum, and J.~L. S\o{}rensen.
\newblock Nondestructive identification of cold and extremely localized single
  molecular ions.
\newblock {\em Phys. Rev. Lett.}, 93(24):243201, Dec 2004.

\bibitem{mor01}
G.~Morigi and H.~Walther.
\newblock Two-species coulomb chains for quantum information.
\newblock {\em European Physical Journal D}, 13:261--269, 2001.

\bibitem{bar03}
M.~D. Barrett, B.~DeMarco, T.~Schaetz, V.~Meyer, D.~Leibfried, J.~Britton,
  J.~Chiaverini, W.~M. Itano, B.~Jelenkovi\ifmmode~\acute{c}\else \'{c}\fi{},
  J.~D. Jost, C.~Langer, T.~Rosenband, and D.~J. Wineland.
\newblock Sympathetic cooling of {$^{9}Be^{+}$ and $^{24}Mg^{+}$} for quantum
  logic.
\newblock {\em Phys. Rev. A}, 68(4):042302, Oct 2003.

\bibitem{sch05}
P.~O. Schmidt, T.~Rosenband, C.~Langer, W.~M. Itano, J.~C. Bergquist, and D.~J.
  Wineland.
\newblock Spectroscopy using quantum logic.
\newblock {\em Science}, 309(5735):749--752, 2005.

\bibitem{ram28}
C.~V. Raman and K.~S. Krishnan.
\newblock A new type of secondary radiation.
\newblock {\em Nature}, 12(3048):501--502, 1928.

\bibitem{ye05}
J.~Ye and S.~T. Cundiff.
\newblock {\em Femtosecond Optical Frequency Comb Technology}.
\newblock Springer, New York, first edition, 2005.

\bibitem{wil80}
E.~B. Wilson, J.~C. Decius, and P.~C. Cross.
\newblock {\em Molecular Vibrations}.
\newblock Dover, New York, first edition, 1980.

\bibitem{hay10}
D.~Hayes, D.~N. Matsukevich, P.~Maunz, D.~Hucul, Q.~Quraishi, S.~Olmschenk,
  W.~Campbell, J.~Mizrahi, C.~Senko, and C.~Monroe.
\newblock Entanglement of atomic qubits using an optical frequency comb.
\newblock {\em Phys. Rev. Lett.}, 104(14):140501--1--4, Apr 2010.

\bibitem{win03}
D.~J. Wineland, M.~Barrett, J.~Britton, J.and~Chiaverini, B.~DeMarco, W.~M.
  Itano, Jelenkovic B., Langer C., D.~Leibfried, V~Meyer, T.~Rosenband, and
  T.~Schaetz.
\newblock Quantum information processing with trapped ions.
\newblock {\em Phil. Trans. R. Soc. Lond. A}, 361:1349--1361, 2003.

\bibitem{vog06}
I.~S. Vogelius, L.~B. Madsen, and M.~Drewsen.
\newblock Probabilistic state preparation of a single molecular ion by
  projection measurement.
\newblock {\em J. Phys. B:At. Mol. Opt. Phys.}, 39:S1259--S1265, 2006.

\bibitem{ber98}
K.~Bergmann, H.~Theuer, and B.~W. Shore.
\newblock Coherent population transfer among quantum states of atoms and
  molecules.
\newblock {\em Rev. Mod. Phys.}, 70(3):1003--1025, Jul 1998.

\bibitem{hum07}
D.~B. Hume, T.~Rosenband, and D.~J. Wineland.
\newblock High-fidelity adaptive qubit detection through repetitive quantum
  nondemolition measurements.
\newblock {\em Phys. Rev. Lett.}, 99(12):120502, Sep 2007.

\bibitem{jor05}
S.~{J{\o}rgenson}, M.~Drewsen, and R.~Kosloff.
\newblock Intensity and wavelength control of a single molecule reaction:
  Simulation of photodissociation of cold-trapped {MgH$^+$}.
\newblock {\em The Journal of Chemical Physics}, 123:094302--1--9, 2005.

\bibitem{din11}
S.~Ding and D.~N. Matsukevich.
\newblock Quantum logic for control and manipulation of molecular ions using a
  frequency comb.
\newblock {\em arXiv:1109.4251v1}, 2011.

\end{thebibliography}
\end{document}